\documentclass[aps,pre,twocolumn,groupedaddress]{revtex4-1}

\usepackage{epsfig}
\newcounter{fofo}      
\begin{document}

\bibliographystyle{unsrt}


\title{\bf Numerical simulation of copper ablation by ultrashort laser pulses}


\author{PengJi Ding, BiTao Hu, YuHong Li*}
\affiliation{School of Nuclear Science and Technology, Lanzhou University, Lanzhou, China}


\date{\today}

\begin{abstract}
    Using a modified self-consistent one-dimensional hydrodynamic lagrangian fluid code, laser ablation of solid copper by ultrashort laser pulses in vacuum was simulated to study fundamental mechanisms and to provide a guide for drilling periodic microholes or microgratings on the metal surface. The simulated laser ablation threshold is a approximate constancy in femtosecond regime and increases as the square root of pulse duration in picosecond regime. The ablation depth as a function of pulse duration shows four different regimes and a minimum for a pulse duration of $\sim 12$ $ps$ for various laser fluences. The influence of laser-induced plasma shielding on ablation depth is also studied. 
\end{abstract}


\maketitle

\section{Introduction}
    Micromachining with ultrashort laser pulses has been a rapidly advancing area of ultrashort laser applications in recent twenty years. For its high precise and practically melting free processing of nearly all kind of materials, it has been applied to many fields such as pulsed laser deposition (PLD), nanoparticle manufacturing, surgery and chemical analysis. It utilizes the properties of femtosecond or picosecond laser pulses to achieve an unprecedented degree of control in sculpting the desired microstructures internal or external to the material without much collateral damage to the surroundings \cite{ablation-01}. The quality of ablated holes and patterns produced with ultrashort laser pulses is much better than the quality of structures produced with nanosecond or microsecond laser pulses. Another new application of ultrashort laser pulses was reported that femtosecond laser processing could create a variety of colors on metal surface by laser-induced periodic microholes and could turned aluminum-like metals into golden appearance \cite{Appl.Phys.Lett.92.041914}. This metal colorization techinique will ultimately allow people to control the optional properties of metal surface.
    
    To control and optimize the process of laser micromachining, the understanding of the fundamental mechanisms of ultrashort laser ablation is required. The coupling of the ultrashort laser pulse to the metals is a complex process. Several theoretical models \cite{ablation-01, simulation-01, theory-01} have been proposed to simulate ultrashort laser ablation of metals. Some other models such as electrostatic ablation \cite{Phys.Plasmas.9.949} and Coulomb explosion \cite{PhysRevLett.88.097603} were also proposed but these models are not supposed to apply for the metals due to their high electron conductivity and fast thermal diffusion. During ultrashort laser ablation of aluminum, as described in Ref. \cite{PhysRevLett.86.2573}, clear separation between the ablated matter and the unablated target has been seen to occur through spinodal decomposition involving thermodynamic instabilities near the critical point of aluminum. Fast heating and phase transition leading to phase explosion seems to be a more realistic mechanism in metal ablation by ultrashort laser pulses and hydrodynamic models can provide a consistent description of the processes of laser ablation of the metals when the laser intensity is not too high but appropriate for micromaching use.  
   
    In this paper, we simulated the copper ablation by femtosecond and picosecond laser pulses with a modified hydrodynamic code MED103 \cite{med103-01,med103-02}. The dependence of the copper laser ablation as a function of laser pulse parameters (such as laser fluence, pulse duration, the number of pulses and repetition rate) and other effects which were involved in the process of copper laser ablation. With the aim to improve our understanding of the physical processes involved in laser-driven ablation and determining the optimum physical condition for manufacturing required surface structures, we have undertaken the development of a computer code treating both the processes of laser-driven ablation and subsequent plasma expansion in vacuum. A one-dimensional fluid model was employed in the computer code which include a self-consistent treatment of hydrodynamics, laser energy absorption, thermal conduction, equilibrium between electrons and ions temperature, fluid motion, shock waves generation and etc. Similar codes \cite{PhysRevE.62.1202, simulation-01, simulation-02} have been developed to simulate ultrashort laser ablation, laser-driven thermonuclear fusion, hot-electron generation and transport, laser-induced plasma spectroscopy in recent twenty years. A set of experimental conditions have to be carefully chosen in order to create required structures on the metal surface, so complete simulation is essential to the success of the experience.
    
    The following section describes the model used in the hydrodynamic code, and Sec.\setcounter{fofo}{3} \Roman{fofo} presents the simulation results. A summary and discussion are given in Sec.\setcounter{fofo}{4} \Roman{fofo}.

\section{Theoretical Model}
    Hydrodynamic codes are a useful tool for a self-consistent description of laser ablation processes when the laser intensity is not too high. At high laser intensities collisionless and relativistic effects become dominant \cite{PhysRevE.66.066415, simulation-01, simulation-02, PhysRevE.62.1202, PhysRevLett.86.2573}. Particle in cell (PIC) simulations including collisions indicate the validity energy range of hydrodynamic simulations goes up to $ S_L\lambda^2 \approx 10^{17} (W/cm^2)\mu m^2 $ where $S_{L}$ is the laser pulse intensity and $\lambda $ is the laser wavelength \cite{PhysRevE.62.1202}. 
    
    The coupling of the ultrashort laser pulses to the metal target is a complicated process which is very different from those of semi-conductors or insulators because of its high thermal conductivities and its low melting temperature. The process involves the absorption of the laser pulse energy in the cold metal surface, the generation of high temperature and hydrodynamic pressure gradients, the propagation of electron heat wave propagating into the bulk of the metal and the laser-induced plasma expansion etc. For femtosecond laser pulses, laser energy is very rapidly absorbed by the electrons meanwhile lattices remain in original temperature. In metals, the conduction band is partially occupied by electrons. The electrons gain energy by inverse-Bremsstrahlung absorption and nonlinear absorptions like photoionization and avalanche ionization to be excited to a higher energy level in the conduction band. At the end of the laser pulse the energy distribution of electrons may be noneqilibrium and this nonequilibrium energy distribution ralax to a Fermi-Dirac distribution in sub-picosecond which can be properly characterized by a macroscopic quantity, i.e, temperature $T$. For picosecond laser pulses, the electron energy distribution generally reaches an equilibrium state at the end of the laser pulses. During the laser ablation thermal process does not have much contribution because the duration of the whole process is shorter than the electron-lattice relaxation time which is typically several picoseconds. As a consequence of rapid laser energy deposition, irradiating metals with ultrashort laser pulses results in strong noneqilibrium conditions between the electrons and the lattice. In this case electrons and lattice subsystem can be described by the classical two-temperature model (TTM) proposed by Anisimov \textit{et al.} \cite{TTM}. The huge temperature gradients from nonequilibrium hot electrons to relatively cold lattices in a constant volume lead to the buildup of high compressive stresses which result in the development of a tensile component of the pressure wave that propagates deeper into the bulk of the target. The tensile stress increases with the depth until it reach the maximum value. If this stress could eventually overcome the mechanical strength of the target material, it could lead to the mechanical separation and ejection of material, i.e., ablation occurs. Electrons and ions temperature eventually reach an equilibrium and then thermal expansion occurs. The metal surface layer on the order of the optical penetration depth is ablated by electron emission, sublimation and transition to the plasma state. The remaining heat diffuses into the metal and leads to the emission of particles and droplets due to the thermal boiling process. The material ejected could carry away most of the deposited laser energy, especially when working very close to the ablation threshold. Because of minimizing the amount of thermal energy diffused into the metal, ultrashort laser ablation greatly reduced the collateral damage to material surrounding the ablation zone.
    
    The code used here is a hydrodynamic Lagrangian fluid code, which includes ionization, collisional and resonance absorption, thermal transport, equation of state (EOS), hydrodynamics and a number of optional physics such as soft X-ray emission and lasing \cite{med103-01, med103-02}. In the hydrodynamic model, the fluid equations for the conservation of mass, momentum, and internal energy are solved with finite difference method. This model treats the electrons and ions as different subsystem characterized by internal energy U, temperature T, pressure  p, specific heat ratio $\gamma$  and so on. They are coupled together by the energy exchange due to the electron-phonon or the electron-atom collisions. In the Lagrangian numerical scheme, the mass of each mesh is constant in time. The velocity $u$ which defines the motion of the Lagrangian coordinates is obtained by solving the Navier-Stokes equation. The temperature is obtained by solving the energy conservation equation:
\begin{equation}
\left(  \frac{\partial U}{\partial T}\right) \frac{dT}{dt}+\left[  \left(  \frac{\partial U}{\partial \rho }\right) _T-\frac{p}{\rho ^2}\right]   \frac{d\rho }{dt}=S
\end{equation} 
  where $U$ is the internal energies per unit mass and source term $S$ is the rate per unit mass at which energy enters each subsystem. The source term $S_i$ and $S_e$ for ions and electrons respectively are wirtten as $ S_i=H_i-K+Q $ and $ S_e=H_e+K+J+X $ where $H$ represents the flow of heat due to thermal conduction, $K=\nabla \cdot \kappa $\textbf{$\nabla$}$ T/\rho $ is the rate of energy exchange between ions and electrons, $J$ is the rate of Bremsstrahlung emission, $X$ is t                                                                                                                                                                                                                                                                                                                                                                                                                                                                                                                                                                                                                                                                                                                                                                                                                                                                                                                                                                                                                                                                                                                                                                                                                                                                                                                                                                                                                                                                                                                                                                                                                                                                                                                                                                                                                                                                                                                                                                                                                                                                                                                                                                                                                                                                                                                                                                                                         he rate of absorption of laser pules and $Q$ is the rate of viscous shock heating. The thermal conductivity $\kappa $ is obtained from the model of L.Spitzer with correction factors. For the hot plasma we use Spitzer's formula for the collision frequency \cite{Spitzer}:
\begin{equation}
\omega _{Spitzer}=\frac{4}{3}(2\pi )^{1/2}\frac{Z^{*}e^{4}m_{e}n_{e}}{(m_{e}\kappa _{B}\textit{T}_{e})^{3/2}}ln(\Lambda )
\end{equation}  
where $Z^{*}$ is the ionization degree and $ln(\Lambda )$ is the Coulomb logarithm. The non-LTE (non-local temperature equilibrium) time-dependent average model was used to calculate the electron energy obtained in the atomic processes of excitation, ionization etc. With this model, the rapidly heated plasmas induced by the ultrashort laser pulse is simulated. The electron thermal heat conductivity $\kappa _{Spitzer} $ is defined by $ \kappa^{*} n_{e} k_{B} T_{e}/(m_{e} \omega _{Spitzer})$, where $\kappa ^{*}$ is given by $\kappa ^{*}= \delta \epsilon 320/(3\pi )$ where $\delta $ and $\epsilon $ are the correction factors, which are weakly dependent on $Z^{*}$ and tabulated \cite{Spitzer}. The exchange rate $K$ of energy between ions and electrons is given by $K = \nu  n_{e} k_{B}(T_{i}-T_{e})/\rho $, where $\nu  $ is the exchange rate expressed as $3m_{e}k_{B}\omega _{Spitzer}/m_{i}^{2}$. Laser absorption is assumed to occur via inverse-Bremsstrahlung at densities below the critical densities $\rho _{c} = \varepsilon _{0} M m_{H} m_{e} \omega _{L}^{2}/(Z e^{2})$ $kg/m^{3}$ at which the laser-induced plasma frequency equals the frequency $\omega _{L}$ of the laser pulse. The resonance absorption is simply considered that twenty-five percent of laser energy having reached critical density is deposited in the plasma by resonance absorption. The distribution of the fast electrons produced by resonance absorption is approximately a Maxwellian distribution at temperature $T_{H}$. The compilation of $T_{H}$ scales as $(\Phi \lambda ^{2})^{1/4}$ for $ (\Phi \lambda ^{2}) > 10^{15}$ $ W cm^{-2} \mu m^{2}$ and as $(\Phi \lambda ^{2})^{2/3}$ for $ (\Phi \lambda ^{2}) < 10^{15}$ $ W cm^{-2} \mu m^{2}$ where $\Phi $ is the laser intensity arriving at the critical density surface. The laser pulses energy absorption term $X(r,t)$ is given by $X(r,t)=P_L(r,t)/dM$, where $P_L(r,t)$ is obtained from laer energy absorption via inverse-Bremsstrahlung at densities below the critical density $\rho _c$ and resonance absorption. The EOS used in the code assumes the ions behaved as a non-degenerate perfect gas and the electrons behaved as a perfect gas which might either be non-degenerate, partially degenerare or fully degenerate. The shock wave created by the expanding plasma is consistently treated as for the target. This model also includes a high-field corrections to inverse-Bremsstrahlung absorption, tunnel ionization, a non-local electron heat conduction and the modelling of collisional in addition to recombination X-ray lasers.

\section{Results}
\subsection{Ablation threshold}
    The most popular industrial application of the ultrashort laser is micromaching such as drilling or cutting. Such a process requires pulse energy enough to give us laser fluence above ablation threshold. However, an interesting application called laser induced periodic surface structure (LIPSS) \cite{PhysRevLett.49.1955} applied for colorizing metals happens near the ablation threshold while even below the ablation threshold the ultrashort laser pulses can induce amorphization on the target without ablation, so the precise value of laser ablation threshold is needed. Ultrashort laser ablation occurs when the laser fluence exceeds a certain threshold fluence which relies on the material properties, laser wavelength, pulse intensity and duration. The target is assumed to be ablated when a rapid decrease in the copper density profile appears. The ablation threshold was determined by the laser fluence when the copper surface density is changed rapidly and the simulated ablation depth exceeded $10$ $nm$ which was estimated to be the minimum ablation depth. Fig. 1 shows the simulation results of the threshold fluence of ablation in copper sample by ultrashort laser pulse as a function of the pulse duration. The threshold fluence increases as laser pulse duration and shows two different regimes with a transition occuring between $1$ $ps$ and $5$ $ps$. For femtosecond pulse, the threshold fluence is approximately a constant value of $0.237$ $J/cm^{2}$, while for pulse longer than $\sim 5$ $ps$ the threshold fluence could be represented as $0.184\tau _L^{1/2}$. The simulated threshold fluence is in qualitative agreement with the simulation results of Ref. \cite{PhysRevE.66.066415, ablation-02} and the experimental data obtained for gold \cite{theory-01}, fused silica and calcium fluoride \cite{PhysRevLett.74.2248}. The constancy of the threshold fluence for femtosecond laser pulse is in good agreement with the measurements made for copper sample \cite{ApplPhysA:MaterSciProcess.61.33}. The constant value of the ablation threshold for femtosecond laser pulse is basically due to the fact that thermal penetration depth $l={D\tau _a}^{1/2}$ is smaller than optical penetration depth $\delta$ where $D$ is the thermal conductivity and $\tau _a$ is the ablation time. In this case threshold fluence $F_{th}$ could be written as $F_{th}\simeq \rho \Omega \delta $ where $\rho $ is the density, $\Omega $ is the specific heat of evaporation. The optical penetration depth $\delta $ can be expressed as $\delta = c/{\omega k}$, where $k$ is the imaginary part of the refractive index. The optical penetration depth does not rely on pulse duration, so the threshold fluences are approximately the same for different pulse durations. The behavior of threshold fluence varying as the square root of the laser pulse duration $\tau_L $ larger than $1$ $ps$ results from the threshold for ablation being controlled primarily by the thermal diffusion of the incident laser pulse enenrgy \cite{PhysRevE.66.066415, theory-01}. In this situation the heat penetration depth is much larger than optical penetration depth and threshold fluence can be written as $F_{th} = \rho \Omega l = \rho \Omega {D\tau _a}^{1/2}$ which is in direct proportion to ${\tau _a}^{1/2}$. Note that in this case, $\tau_L \gg \tau _i$, the ablation time $\tau _a\approx \tau _e \tau _L/(\tau _e+\tau _i)$, where $\tau _e$ and $\tau _i$ are the electron cooling and the lattice heating times, so threshold fluence varies as $\tau_L^{1/2}$ which is understandable.
\begin{figure}[h]
   \begin{center}
     \includegraphics[width=.5\textwidth,angle=0]{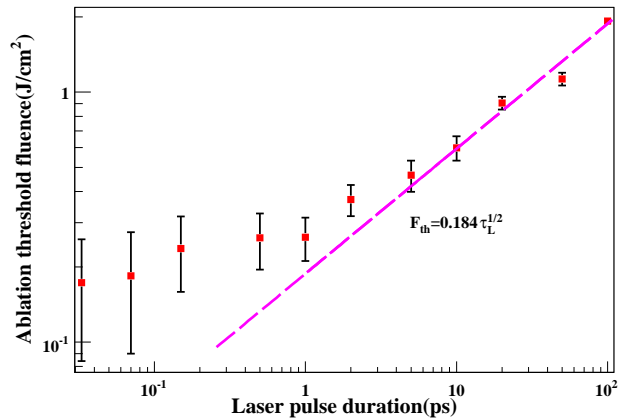}
    \end{center}
    \caption{Threshold fluence vs pulse durations for copper ablation in vacuum ($800$ $nm$ wavelength, $33 fs-100 $ $ps$ pulse duration range).}
    \label{fig:compressed:eps}
\end{figure}

\subsection{Ablation depth}
   From theoretical analysis derived from one-dimensional two-temperature model \cite{TTM}, the ablation depths can be expressed as
\begin{equation}
L\simeq \delta \texttt{ln}(F/F_{th}^{\delta })   \mbox{                               }(\delta \gg l).
\end{equation}
\begin{equation}
L\simeq l \texttt{ln}(F/F_{th}^{l})   \mbox{                               }(\delta \ll l).
\end{equation}
Fig. $2$ shows the ablation depth of copper as a function of the laser fluences obtained by our simulation with fixed pulse duration $\tau _{L}=150$ $fs$ and laser wavelength $\lambda =780$ $nm$. The ablation depth generally shows two different logarithmic dependence on the laser fluence which is in qualitative agreement with the reported experimental results~\cite{ablation-01}. For laser fluence $F$ less than $0.9$ $J/cm^{2}$, a good fit to the simulation results with equation $(3)$ gives the threshold fluence $F_{th}^{\delta } \simeq 31 $ $mJ/cm^{2}$ and optical penetration depth $\delta \simeq 10.4$ $nm$ which is in good agreement with standard value $\delta = 13$ $nm$ \cite{book-01}. In this regime, the laser energy absorption coefficient is so low that not much energy transferred  deeper than the skin depth of the copper target. This ablation regime is dominated by the optical penetration and the influence of thermal conduction can be negligible. One observes a transition occurring between $0.9$ $J/cm^{2}$ and $2.0$ $J/cm^{2}$. In this regime, the influence of electronic heat diffusion becomes more obvious with the increase of laser fluence. At higher fluence ($F > 2.0$ $J/cm^{2}$), the simulation results fitted with equation $(4)$ gives the threshold fluence $F_{th}^{l} \simeq 1.470$ $J/cm^{2}$ and thermal penetration depth $l \simeq 126.8$ $nm$. The obtained thermal penetration depth is bigger than experimental results \cite{ablation-01} because in that experiment the laser pulses are influenced by the walls induced by former pulses which decreased the effective penetration depth.    
\begin{figure}[h]
   \begin{center}
    \includegraphics[width=.5\textwidth,angle=0]{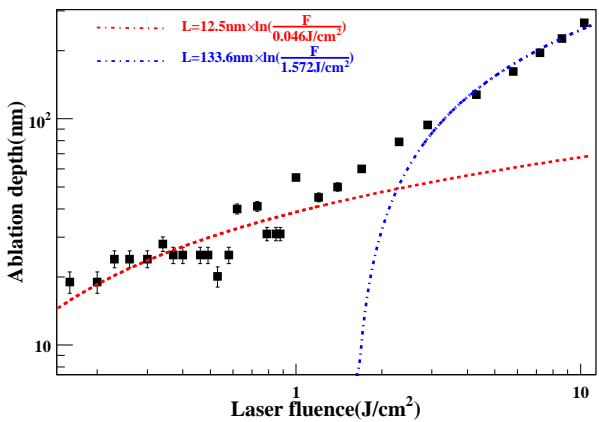}
    \end{center}
    \caption{Copper ablation depth versus laser fluence for 150fs laser pulse. The dash and the dash-dot lines represent the most compatible fits, where $F_{th}^{\delta } = 0.031$ $J/cm^{2}$ and $F_{th}^{l} = 1.470$ $J/cm^{2}$.}
    \label{fig:compressed:eps}
\end{figure} 

   In Fig. 3, simulation results of the ablation depth for different pulse durations from $500$ $fs$ to $20$ $ps$ are summarized. There have no much variations between the ablation depths in case of $500$ $fs$ and $1$ $ps$ where the laser energy deposition is determined by the optical penetration depth as explained before. In case of pulse duration $\tau =3$ $ps$ and $\tau =5$ $ps$, the ablation depth values are obviously lower than the ones of $500$ $fs$ and $1$ $ps$ for same laser fluence in the regime $F > 1$ $J/cm^{2}$. In this regime, the logarithmic dependence on the laser fluence still works, but the ablation depths and the threshold fluences are now dependent on the pulse duration. Increasing pulse duration reduces the ablation depth for the same laser fluence. In case of pulse duration longer than $10$ $ps$, no logarithmic dependence on the laser fluence can be observed in the region $F > 2$ $J/cm^{2}$. As described in Ref. \cite{ablation-01}, this effect can be explained by the hydrodynamic plasma expansion during the laser pulse, plasma shielding of the laser radiation, and increased heat-conduction losses.
\begin{figure}[h]
   \begin{center}
    \includegraphics[width=.5\textwidth,angle=0]{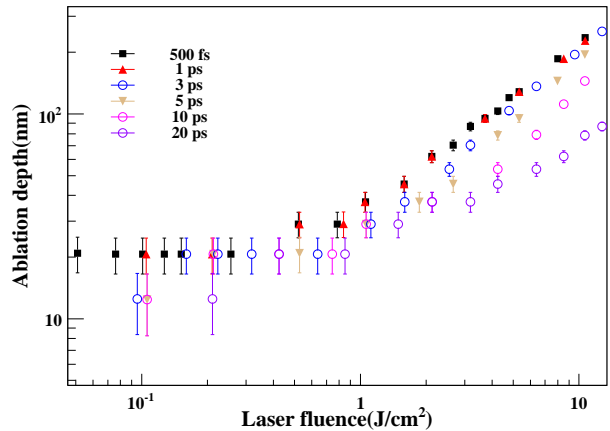}
    \end{center}
    \caption{Copper ablation depth versus laser fluence for different pulse duration: $500$ $fs$, $1$ $ps$, $3$ $ps$, $5$ $ps$, $10$ $ps$, $20$ $ps$.}
    \label{fig:compressed:eps}
\end{figure}
 
   Fig. $4$ presents the simulation results of the copper ablation depth versus the pulse duration $\tau $ with different laser fluences. The ablation depth decrease rapidly in the region $\tau < 150$ $fs$. In this region resonance absorption plays an important role in the process of laser energy absorption and produces hot-electrons with huge energy penetrating deeply into the bulk of the copper target. In this simulation situation the temperature of hot-electron is given by $T_{H}\sim (\Phi  \lambda ^{2}/\tau )^{2/3}$. With the same laser fluence, the bigger pulse duration $\tau $ is, the smaller the hot-electron's temperature is which is equivalent to the smaller ablation depth.
\begin{figure}[h]
   \begin{center}
    \includegraphics[width=.5\textwidth,angle=0]{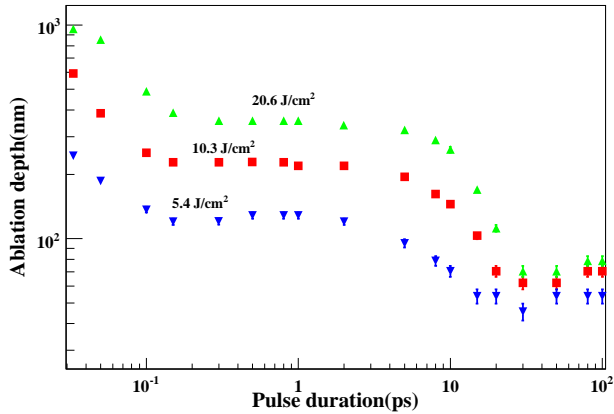}
    \end{center}
    \caption{Ablation depth as a function of the pulse duration for different laser fluences of 5.4, 10.3, and 20.6 $J/cm^{2}$.}
    \label{fig:compressed:eps}
\end{figure} 
No significant change in ablation depth is observed in the range of laser pulse duration from $0.15$ $ps$ to $2$ $ps$. The relative constancy in this region can be explained by the arguments used in the ablation threshold subsection. In the region $2$ $ps$ $< \tau < 50$ $ps$, the ablation depth decreases with the increase of the pulse duration. In this $ps$ regime, the laser energy absorption is dominated by inverse-Bremsstrahlung, which results in a decrease in the amount of energy reaching the copper surface, so that the ablation depth decreases as well. One observes a minimum in the ablation depth occurs in region $10< \tau < 100$ $ps$. In the region $50$ $ps$ $< \tau < 100$ $ps$ the ablation depth increases a little up to $70$ $nm$ at first and then reaches a relative constant value. An analytical investigations in Ref. \cite{theory-02} indicated that the laser absorption increased with $\Lambda / \lambda $ where $\Lambda = {|\partial_{z} {ln(n_{e})}|} _{n_{e}=n_{c}}^{-1}$ is the electron density gradient length. The ablation depth is proportion to the total absorbed laser energy in case of Gaussian pulses, so the increase of the ablation depth in this region can be explained by the the increase of the energy absorption in the laser-induced electron plasma. 

\subsection{Multiplepulse}
   Laser ablation efficiency can be optimized by using a prepulse to create a plasma profile on the surface of the metal before the main pulse \cite{simulation-01}. This effect is determined by the timing between the prepulse and the main pulse because if the temporal separation is too long, laser energy is absorbed in the plasma far away from the solid surface and does not contribute to the laser ablation. In this situation, ablation depth largely depends on the lag between the prepulse and the main pulse. The simulation results of ablation depth as a function of the lag between the prepulse and the main pulse are shown in Fig. $5$. The general shape of the curves is in qualitative agreement with the experimental results described in Ref. \cite{ablation-03}. Three different regimes of the laser ablation denoted with Latin alphabets depending on the lag between the prepulse and the main pulse. The constancy of the ablation depth in region \textbf{A} where the lag is less than $1.3$ $ps$ indicates that the double pulses are close enough to each other which could be considered as one pulse. With the lag longer than $8$ $ps$, in region \textbf{C}, the main pulse is completely shielded by the prepulse-induced plasma. The region \textbf{B} of $1.3$ $ps$ $\sim 8$ $ps$ corresponds to the transition regime with partial plasma shielding.
\begin{figure}[h]
   \begin{center}
    \includegraphics[width=.5\textwidth,angle=0]{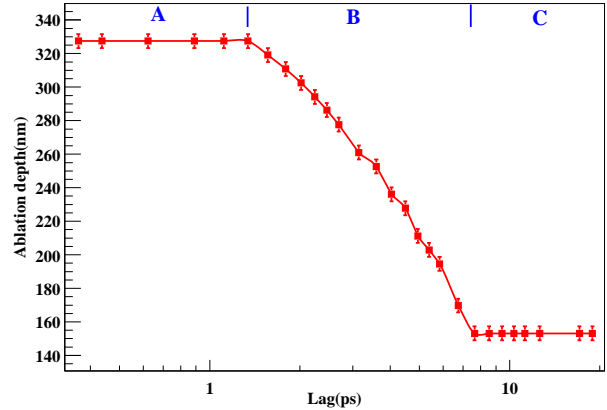}
    \end{center}
    \caption{Copper ablation depth as a function of the lag between main pulse and prepulse. The used prepulse and main pulse parameters: pulse duration $\tau = 150$ $fs$, laser wavelength $\lambda =1.0\mu$ $m$. The fluence of prepulse is about $1.64$ $J/cm^{2}$ and total laser fluence is about $18$ $J/cm^{2}$.}
    \label{fig:compressed:eps}
\end{figure}
    
   We utilize upto five laser pulses to simulate the dependence of the ablation depth on laser pulse number in order to investigate the influence of plasma shielding on copper ablation in case of several laser pulses. Fig. $6$ indicates the total ablation depth increase with increasing number of laser pulses, while ablation depth per pulse decrease with increasing number of pulses. The delay between two pulses was set as to be near $3.0$ $ps$ and the fluence of each pulse was set to be near $8$ $J/cm^{2}$. Experimental results presented in Ref. \cite{ablation-04} indicate an approximate linear increase of ablation depth with increasing number of laser pulses. Our simulation results deviate from the linear dependence apparently. The reason is that the lag between pulses in our simulation is so short that the influence of plasma shieding is important, while in the experiment the lag between pulses was so long that the plasma shielding effect can be neglect and preceding pulses almost have no influence on subsequent pulses. Ultrashort laser drilled holes typically exhibit a conical shape due to the Gaussian intensity profile of the laser beam and enhanced thermal losses at the walls of the hole. As a consequence, the laser pulses are interacting with the walls at grazing incidence increasing the interaction area that decrease the effective fluence.
\begin{figure}[h]
   \begin{center}
    \includegraphics[width=.5\textwidth,angle=0]{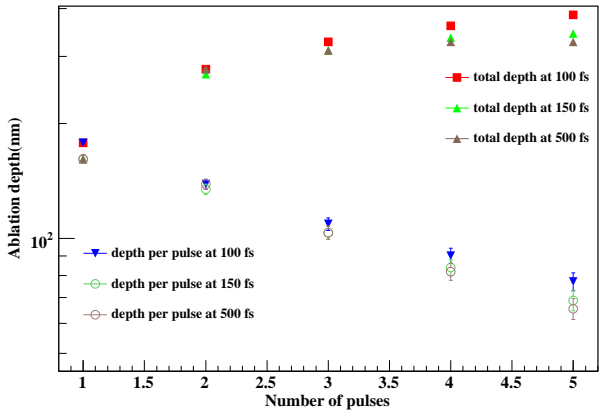}
    \end{center}
    \caption{Total ablation depth and ablation depth per pulse as a function of the number of pulses in different pulse duration $100$ $fs$, $150$ $fs$, $500$ $fs$. }
    \label{fig:compressed:eps}
\end{figure}

\section{Summary}
    This paper presents the simulation results and physical interpretations of copper ablation by ultrashort laser. The ablation threshold shows a approximate constant value for pulse duration less than $1$ $ps$ and increases as $0.184\tau_{L} ^{1/2}$ for pulse duration $\tau_{L} > 2$ $ps$ after a transition occuring between $1$ $ps$ and $2$ $ps$. This dependency relationship can be analytically explained by different ablation processes in the cases of pulse duration smaller or greater than the electron-lattice relaxaion time. In the former case, deposited laser energy during the pulse duration has no much difference compared to laser energy leading to ablation after the complete of equilibrium between the electron temperature and the ion temperature. After electron-lattice ralaxaion time, the ablation process is dominated by thermal diffusion in which the ablation threshold fluence varies as $\tau _{L}^{1/2}$. The two logarithmic dependence on the laser fluence of ablaion depth is simulated which is in good agreement with experimental results reported in Ref. \cite{ablation-01}. The dependence of ablation depth on pulse duration for a certain laser fluence clearly shows the efficiency of laser ablation of copper in femtosecond regime is higher compared to the one in picosecond regime. A experimental investigation of Ref. \cite{PhysRevE.56.7179} indicate that laser ablation efficiency can be optimized by using a prepulse to create a preplasma for the main pulse, but our simulation of copper laser ablation shows a continuing reduction of ablation depth as the increase of the lag between the prepulse and main pulse. Because of the plasma shielding effect, the laser energy of subsequent pulses can't completely contribute to the laser ablation of copper if the lag between two pulses is too short. In general we should set the lag between two pulses larger than dozens of nanoseconds when the effect of plasma shielding has no much influence.

\begin{acknowledgments}
  We gratefully acknowledge the support of the laser staff at Laboratory of Laser Nuclear Physics. This work was supported by the Program for New Century excellent Talents in University, National Natural Science Foundation of China (Grant No.10775062, 10875054 and 10975065) and the Fundamental Research Funds for the Central Universities (Grant No. lzujbky-2010-k08).
\end{acknowledgments}
\bibliography{myreference}

\end{document}